\documentclass[superscriptaddress,aps,11pt,nofootinbib]{revtex4}
\usepackage{graphicx}
\usepackage{amsmath}
\usepackage{bm}

\begin{document}

\preprint{}
\title{PHENOMENOLOGICAL AND COSMOLOGICAL ASPECTS OF A MINIMAL GUT SCENARIO}
\author{Ilja Dorsner}
\email{idorsner@ictp.trieste.it} \affiliation{The Abdus Salam
International Centre for Theoretical Physics\\
Strada Costiera 11, 34014 Trieste, Italy}
\author{Pavel Fileviez P\'erez}
\email{fileviez@cftp.ist.utl.pt} \affiliation{Departamento de F\'{\i}sica and Centro
de F\'{\i}sica Te\'orica de Part\'{\i}culas, Instituto Superior T\'{e}cnico, Av. Rovisco
Pais, 1049-001 Lisboa, Portugal}
\author{Ricardo Gonz\'alez Felipe}
\email{gonzalez@cftp.ist.utl.pt} \affiliation{Departamento de F\'{\i}sica and Centro
de F\'{\i}sica Te\'orica de Part\'{\i}culas, Instituto Superior T\'{e}cnico, Av. Rovisco
Pais, 1049-001 Lisboa, Portugal}

\begin{abstract}
Several phenomenological and cosmological aspects of a minimal extension of the
Georgi-Glashow model, where the Higgs sector is composed by $\bm{5}_H$,
$\bm{15}_H$, and $\bm{24}_H$, are studied. It is shown that the constraints
coming from the unification of gauge interactions up to two-loop level predict
light scalar leptoquarks. In this GUT scenario, the upper bound on the total
proton decay lifetime is $\tau_p \leq 1.4 \times 10^{36}$ years. The
possibility to explain the matter-antimatter asymmetry in the universe through
the decays of $SU(2)_L$ scalar triplets is also studied. We find that a
successful triplet seesaw leptogenesis implies an upper bound on the scalar
leptoquark mass,  $M_{\Phi_b}
 \lesssim 10^{6-7}$~GeV. We conclude that this GUT scenario can be tested at
the next generation of proton decay experiments and future colliders through
the production of scalar leptoquarks.
\end{abstract}
\maketitle
\section{Introduction}
Grand unified theories (GUTs) based on the $SO(10)$ gauge
symmetry~\cite{SO(10)1,Fritzsch:1974nn} are usually considered as the most
appealing candidates for the unification of electroweak and strong
interactions. They offer a number of advantages over $SU(5)$
theories~\cite{GG}: (i) They provide a natural explanation of the smallness of
neutrino masses through the seesaw mechanism~\cite{seesaw}; (ii) they
accommodate all fermions of one generation into one representation; (iii) they
represent, in their minimal form, the most promising theory of fermion masses
(For realistic grand unified theories based on the $SO(10)$ gauge symmetry see
e.g. Refs.~\cite{Babu:1998wi,Aulakh:2003kg,Babu:2005gx}.).

Nevertheless, it is well known that the only promising way to test the idea of
grand unification is through nucleon decay. Therefore, it is very important to
investigate the simplest realistic grand unified theory where proton decay can
be well predicted. This crucial issue brings us back to non-supersymmetric GUT
scenarios, since the unification scale is rather low $(M_{GUT} \approx 10^{14}~
\textrm{GeV})$. In particular, we focus on the simplest realistic $SU(5)$
theories, where the unification scale can be accurately predicted. Even though
$SU(5)$ possesses uncorrelated regions in the Yukawa sector, the simplicity of
the Higgs sector in the non-supersymmetric case offers a hope that the theory
can be verified in near future.

In a recent work~\cite{Dorsner:2005fq}, some of us argued that the simplest
realistic extension of the Georgi-Glashow (GG) model is the one containing the
$\bm{5}_H$, $\bm{15}_H$ and $\bm{24}_H$ representations in the Higgs sector.
The purpose of this paper is to demonstrate that the next generation of
collider and proton decay experiments will refute or verify this minimal
$SU(5)$ scenario. A first attempt was made in Ref.~\cite{Dorsner:2005fq} in
this direction. Here we offer the full two-loop treatment of the gauge coupling
unification and discuss the constraints on the Higgs sector. To show the
testability of the model, we include all the presently available experimental
limits in our discussion. The upper bound on the total proton decay lifetime in
our GUT model is corrected, and we investigate the possibility to explain the
baryon asymmetry observed in the universe through the decays of $SU(2)_L$
scalar triplets living in $\bm{15}_H$, showing how important the constraints
coming from leptogenesis turn out to be. The latter, when combined with the
unification constraints, lead us to conclude that the present GUT scenario
could be tested at the next generation of collider experiments through the
production of light leptoquarks.

\section{A minimal $SU(5)$ scenario}
\label{SU(5)theory}

Ever since its inception in 1974, the $SU(5)$ model of Georgi and
Glashow~\cite{GG} has been considered as the minimal grand unified theory. It
offers partial matter unification of one standard model (SM) family $a$
($a=1,2,3$) in the anti-fundamental ${\bf \overline{5}}_a$ and antisymmetric
${\bf 10}_a$ representations. The GUT symmetry is broken down to the standard
model by the vacuum expectation value (VEV) of the Higgs field in the
$\bm{24}_H$, while the SM Higgs resides in the $\bm{5}_H$. The beauty of the
model is undeniable, but the model itself is not realistic. Indeed, there are
several problems some of which are correlated:

\begin{itemize}

\item Gauge coupling unification.

The most dramatic problem of the naive $SU(5)$ is the lack of unification.
Namely, using the whole freedom of the model, one can compute the maximum value
of the ratio between the parameters $B_{23}=b_2 - b_3$ and $B_{12}=b_1 - b_2$,
where $b_i$ ($i=1,2,3$) are the beta functions of the particle content of the
theory for $U(1)_Y$, $SU(2)_L$ and $SU(3)_C$, respectively. One gets
$B_{23}^{SU(5)}/ B_{12}^{SU(5)} \leq 0.60$, while the present experimental data
requires $B_{23}/B_{12} = 0.719 \pm 0.005$.

\item Relation between Yukawa couplings of quarks and leptons.

A second major problem is related to the predicted relation $Y_D = Y_E^T$
between the down quark and charged lepton Yukawa coupling matrices. This
prediction is in strong disagreement with the experiment, especially in the
case of the first and second generations of fermions. There are two solutions
to this issue: one can add higher-dimensional operators~\cite{Ellis:1979fg} or
introduce the ${\bf 45_H}$ representation~\cite{Georgi:1979df}.

\item Neutrino masses.

In the Georgi-Glashow model, neutrinos are massless. However, today we know
that they do have a tiny mass. Therefore, the model has to be extended in order
to account for non-vanishing neutrino masses. There are two possible solutions:
one can introduce at least two right-handed neutrinos and use the so-called
Type I seesaw mechanism~\cite{seesaw}, or one can add the representation ${\bf
15}_H$ in order to generate neutrino masses through the Type II seesaw
mechanism~\cite{TypeII}.

\item Doublet-triplet (DT) splitting problem.

Another problem in the naive $SU(5)$ is that it cannot explain why the Higgs
doublet living in ${\bf 5}_H$ is light. Although there are no solutions to this
issue in the context of a non-supersymmetric scenario, different mechanisms are
conceivable in SUSY $SU(5)$ to achieve the splitting between the triplet and
the doublet. (See for example Ref.~\cite{Berezhiani:1997as} for a review.)

\end{itemize}

The simplest way~\cite{Dorsner:2005fq} to address the first three problems
listed above in a non-supersymmetric framework consists of extending the
minimal GG model with the ${\bf 15}_H$ and allowing for higher-dimensional
operators. More precisely, the Higgs sector is $\bm{24}_H= \Sigma =(\Sigma_8,
\Sigma_3, \Sigma_{(3,2)}, \Sigma_{(\bar{3}, 2)},
\Sigma_{24})=(\bm{8},\bm{1},0)+(\bm{1},\bm{3},0)+(\bm{3},\bm{2},-5/6)
+(\overline{\bm{3}},\bm{2},5/6)+(\bm{1},\bm{1},0)$, $\bm{15}_H= \Phi =(\Phi_a,
\Phi_b, \Phi_c)= (\bm{1},\bm{3},1)+(\bm{3},\bm{2},1/6)+(\bm{6},\bm{1},-2/3)$,
$\bm{5}_H= \Psi=(\Psi_D, \Psi_T)=(\bm{1},\bm{2},1/2)+(\bm{3},\bm{1},-1/3)$,
where $\Sigma_{(3,2)}$ and $\Sigma_{(\bar{3}, 2)}$ are fields eaten by the
superheavy gauge fields $V$. In what follows we define the GUT scale through
their mass: $M_{GUT}=M_V$. As emphasized in~\cite{Dorsner:2005fq}, in this
non-supersymmetric grand unified model, the GUT scale is low and can be
predicted with great precision. This gives us the possibility to test the grand
unification idea at future proton decay experiments. In fact, in our view,
grand unified theories are the theories for the decay of the proton, since they
provide us with the necessary input to compute the corresponding partial
lifetimes. Of course, one can also think of other minimal extensions of grand
unified theories based on higher groups. However, since in those models it is
very difficult to predict the GUT scale and the masses of the superheavy gauge
bosons mediating nucleon decay, the predictions for the lifetime of the proton
are far from accurate.

In the GUT scenario proposed in~\cite{Dorsner:2005fq} the scalar potential
reads as:
\begin{eqnarray}
V \ & = & \ V_{SU(5)}^{\rm naive} \ + \ \overline{5}^T_{ai} \ C \ h^{ab} \
\overline{5}_{bj} \ 15_H^{ij}  \ - \ \frac{\mu_{\Phi}^2}{2} \text{Tr}\,
15_H^{\dagger} 15_H \ + \ \frac{a_{\Phi}}{4} ( \text{Tr}\, 15_H^{\dagger} 15_H
)^2 \nonumber \\ & + & \frac{b_{\Phi}}{2} \, \text{Tr}\, ( 15_H^{\dagger} 15_H
15_H^{\dagger} 15_H) \ + \ c_2 \, \text{Tr}\, (15_H^{\dagger} 24_H 15_H)  \ + \
c_3 \ 5_H^{\dagger} \ 15_H \ 5_H^* \nonumber \\ &+&  \ c_3^* \ 5_H^T \
15_H^{\dagger} 5_H \ + \ b_1 \, \text{Tr}\,( 15_H^{\dagger} 15_H) \text{Tr}\,
24_H^2 \ + \ b_3 \ 5_H^{\dagger} 5_H \, \text{Tr}\,( 15_H^{\dagger} 15_H) \ \nonumber \\
& + & \ b_{5} \ 5_H^{\dagger} 15_H 15_H^{\dagger} 5_H \ + \ b_6 \, \text{Tr}\,
( 15_H
15_H^* 24_H^2) \ + \ b_7 \, \text{Tr}\, (15_H^* 24_H 15_H 24_H^* ) \nonumber \\
& + & \ b_8 \ 5_H^{\dagger} 24_H 15_H 5_H^* \ + \ b^*_8 \ 5_H^{T} 24_H
15^{\dagger}_H 5_H \ + \ \textrm{higher-dimensional terms},
\end{eqnarray}
where $V_{SU(5)}^{\rm naive}$ is the scalar potential of the Georgi-Glashow
model.

\section{Unification}
\label{unification}

In this section we present the constraints that exact gauge coupling
unification places on the masses of the scalars of the theory. We first present
the one-loop level analysis to outline the basic features of the scalar mass
spectrum and only then we resort to the more accurate two-loop analysis.

\subsection{One-Loop Analysis}

The one-loop level relations between the gauge couplings at $M_Z$ and the
unifying gauge coupling $\alpha_{GUT}=g^2_{GUT}/(4 \pi)$ at $M_{GUT}$ are
\begin{equation}
\label{running} \left.\alpha^{-1}_i\right|_{M_Z}=\alpha^{-1}_{GUT}+\frac{b_i}{2
\pi} \ln \frac{M_{GUT}}{M_Z},
\end{equation}
where $i=1,2,3$ for $U(1)$, $SU(2)$, and $SU(3)$, respectively, and $b_i$ are
the familiar one-loop $\beta$ function coefficients. The SM particle content
with $n$ light Higgs doublet fields yields $b_1=40/10+n/10$, $b_2=-20/6+n/6$
and $b_3=-7$.

Eqs.~\eqref{running} hold under the assumption that there is a particle
``desert'' between $M_Z$ and $M_{GUT}$. However, there is no particular reason
that this should be the case. If there are $I$ particles with intermediate
masses $M_I$ ($M_Z \leq M_I \leq M_{GUT}$), these equations remain unaltered
except for the substitutions $b_i \rightarrow B_i$, where $B_i = b_i+\sum_{I}
b_{iI} r_{I}$ are the so-called effective coefficients. Here $b_{iI}$ are the
appropriate one-loop coefficients of the particle $I$ and $r_I=(\ln
M_{GUT}/M_{I})/(\ln M_{GUT}/M_{Z})$ ($0 \leq r_I \leq 1$) is its ``running
weight''.

The elimination of $\alpha_{GUT}$ from Eqs.~\eqref{running} leaves the
following two equations that connect the effective coefficients
$B_{ij}=B_i-B_j$ with the low-energy observables~\cite{Giveon:1991zm}:
\begin{equation}
\frac{B_{23}}{B_{12}}=\frac{5}{8} \frac{\sin^2
\theta_W-\alpha_{em}/\alpha_s}{3/8-\sin^2 \theta_W}\,, \qquad \ln
\frac{M_{GUT}}{M_Z}=\frac{16 \pi}{5} \frac{3/8-\sin^2 \theta_W}{\alpha_{em}
B_{12}}\,.
\end{equation}
Adopting the following experimental values at $M_Z$ in the $\overline{MS}$
scheme~\cite{Eidelman:2004wy}: $\sin^2 \theta_W=0.23120 \pm 0.00015$,
$\alpha_{em}^{-1}=127.906 \pm 0.019$ and $\alpha_{s}=0.1187 \pm 0.002$, these
read
\begin{subequations}
\label{conditions}
\begin{eqnarray}
\label{condition1}
\frac{B_{23}}{B_{12}}&=&0.719\pm0.005\,,\\
\nonumber\\
\label{condition2} \ln \frac{M_{GUT}}{M_Z}&=&\frac{184.9 \pm 0.2}{B_{12}}\,.
\end{eqnarray}
\end{subequations}
Eq.~\eqref{condition1} is sometimes referred to as the $B$-test. It basically
shows whether unification takes place or not. Eq.~\eqref{condition2}, on the
other hand, could be referred to as the GUT scale relation since it yields the
GUT scale value once Eq.~\eqref{condition1} is satisfied.

The $B$-test fails badly in the SM case ($B_{23}/B_{12}=0.53$), and hence the
need for extra light particles with suitable $B_{ij}$ coefficients to bring the
value of the $B_{23}/B_{12}$ ratio in agreement with its experimental value. In
our case the presence of the $\bm{15}_H$ is essential. The $B_{ij}$
coefficients for all the particles in our scenario are presented in
Table~\ref{tab:table1}. Clearly, $\Sigma_3$, $\Phi_a$ and $\Phi_b$ improve
unification with respect to the SM case, while $\Sigma_8$, $\Psi_T$ and
$\Phi_c$ act in the opposite manner. We recall that we set $M_V=M_{GUT}$, where
$M_V$ is the mass of the superheavy gauge bosons. We thus take $\Sigma_8$,
$\Psi_T$ and $\Phi_c$ to reside at or above the GUT scale in our numerical
analysis. We relax this assumption later to discuss its impact on our findings.
\begin{table}[h]
\caption{\label{tab:table1} $B_{ij}$ coefficients.}
\begin{ruledtabular}
\begin{tabular}{lccccccccc}
     &Higgsless SM&$\Psi_D$&$\Psi_T$ & $V$ & $\Sigma_8$
     & $\Sigma_3$ & $\Phi_a$ & $\Phi_b$ & $\Phi_c$\\
\hline $B_{23}$& $\frac{11}{3}$&$\frac{1}{6}$&$-\frac{1}{6}
r_{\Psi_T}$ &$-\frac{7}{2}r_V$ &$-\frac{1}{2}
r_{\Sigma_8}$&$\frac{1}{3} r_{\Sigma_3}$ &$\frac{2}{3}r_{\Phi_a}$
&$\frac{1}{6} r_{\Phi_b}$ &$-\frac{5}{6} r_{\Phi_c}$\\
$B_{12}$&$\frac{22}{3}$&$-\frac{1}{15}$&$\frac{1}{15} r_{\Psi_T}$
&$-7r_V$ &0 &$-\frac{1}{3} r_{\Sigma_3}$
&$-\frac{1}{15}r_{\Phi_a}$ &$-\frac{7}{15} r_{\Phi_b}$ &$\frac{8}{15} r_{\Phi_c}$\\
\end{tabular}
\end{ruledtabular}
\end{table}

The value of $B_{12}$ determines the scale of unification through the GUT scale
relation~\eqref{condition2}. Therefore, a lower bound on $B_{12}$ translates
into an absolute upper bound on the GUT scale. If we naively set
$M_{\Sigma_3}=M_{\Phi_a}=M_{\Phi_b}=M_Z$
($r_{\Sigma_3}=r_{\Phi_a}=r_{\Phi_b}=1$) we obtain $B_{12}>6.4$ and accordingly
$M_{GUT}<3.2 \times 10^{14}$\,GeV. This constraint, when combined with the
relation $M_{GUT}=M_V$, determines an upper bound on the proton lifetime, if
the corresponding value of $\alpha_{GUT}$ is known. To find the latter we
resort to the numerical analysis. However, the non-supersymmetric nature of our
scenario suggests this value to be around $1/39$.

\begin{figure}[h]
\begin{center}
\includegraphics[width=4in]{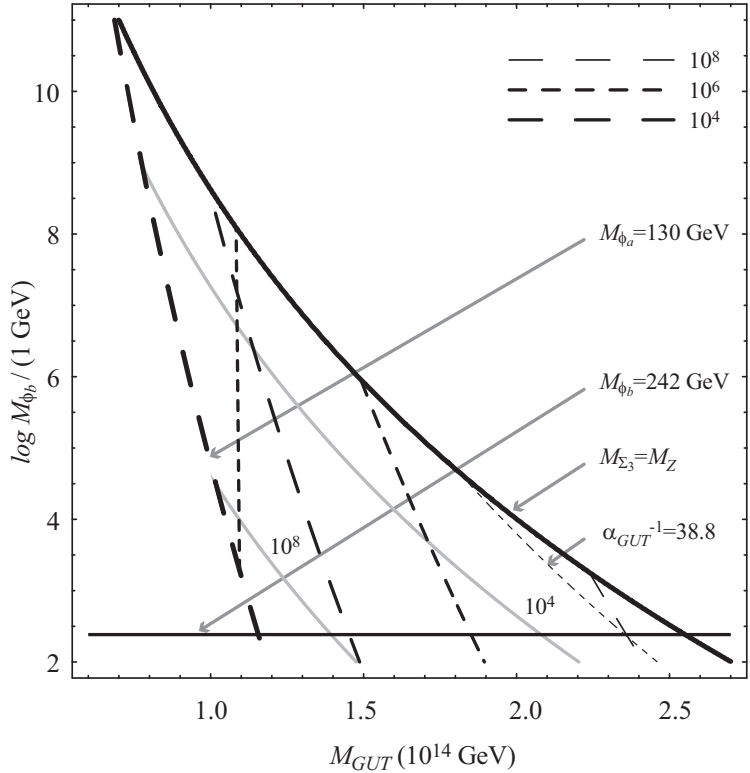}
\end{center}
\caption{\label{figure:1} Plot of lines of constant $M_{\Sigma_3}$ and
$M_{\Phi_a}$ in the $M_{GUT}$--$\textit{log}\, (M_{\Phi_b}/1\,\textrm{GeV})$
plane, assuming exact one-loop unification. We use the central values for the
gauge couplings as given in the text. All the masses are given in GeV units.
The triangular region is bounded from the left (below) by the experimental
limit on $M_{\Phi_a}$ ($M_{\Phi_b}$). The right bound is $M_{\Sigma_3} \geq
M_Z$. The two grey solid (thick dashed) lines are the lines of constant
$M_{\Sigma_3}$ ($M_{\Phi_a}$). The line of constant $\alpha^{-1}_{GUT}\,$ is
also shown. The region to the left of the vertical dashed line is excluded by
the proton decay experiments if $\alpha=0.015$\,GeV$^3$.}
\end{figure}

In the one-loop analysis we treat $M_{GUT}$, $M_{\Sigma_3}$, $M_{\Phi_a}$ and
$M_{\Phi_b}$ as free parameters and investigate the possibility to find a
consistent scenario with \emph{exact} gauge coupling unification. Since we have
four free parameters and two equations---Eqs.~\eqref{condition1}
and~\eqref{condition2}---we opt to present the $M_{\Sigma_3}$ and $M_{\Phi_a}$
contours in the $M_{GUT}$--$M_{\Phi_b}$ plane in Fig.~\ref{figure:1}. The line
of constant $\alpha^{-1}_{GUT}$ is also shown.

The triangular region in Fig.~\ref{figure:1} represents the available parameter
space under the assumption that $\Psi_T$, $\Sigma_8$ and $\Phi_c$ reside at or
above the GUT scale. The region is bounded from the left and below by the
experimental limits on $M_{\Phi_a}$ and $M_{\Phi_b}$, respectively. (For the
discussion on the origin of the experimental limits see~\cite{Dorsner:2005fq}
and references therein.) The right bound stems from the requirement that
$M_{\Sigma_3} \geq M_Z$. It is expected that the Large Hadron Collider (LHC)
will place a more stringent lower limit on the mass of the scalar leptoquark
$\Phi_b$ at around $1$~TeV (For experimental bounds on leptoquark masses see
Refs.~\cite{Leptoexperiments}.).

Fig.~\ref{figure:1} reveals that the masses of the three scalar fields that
improve unification, namely, $\Sigma_3$, $\Phi_a$ and $\Phi_b$, have to be
below the GUT scale. This, however, does not hold at two-loop level. The GUT
scale is rather low and, for a given value of $\alpha_{GUT}$, the predicted
value for the proton decay lifetime is within the reach of the present and
future proton decay experiments. More precisely, if the nucleon matrix element
is taken to be $\alpha=0.015$\,GeV$^3$, the region to the left of the vertical
thick dashed line in Fig.~\ref{figure:1} is already excluded by the present
limits on the proton decay lifetime. In order to generate this bound we assume
maximal flavor suppression of  the gauge $d=6$ proton decay operators as
explained in the next section in more details. Clearly, due to the simplicity
of our scenario, experimental limits place firm upper bounds on $M_{\Sigma_3}$,
$M_{\Phi_a}$ and $M_{\Phi_b}$.

What happens if we relax the $M_{\Psi_T},M_{\Sigma_8},M_{\Phi_c} \geq M_{GUT}$
assumption? If either $M_{\Psi_T}$ or $M_{\Phi_c}$ are below the GUT scale,
then they both decrease $M_{GUT}$ due to an increase of the $B_{12}$
coefficient. They also change the $B_{23}/B_{12}$ ratio in the wrong direction,
which has to be compensated by appropriate changes in the $\Sigma_3$, $\Phi_a$
and $\Phi_b$ contributions. Pictorially, as one lowers $M_{\Psi_T}$ and
$M_{\Phi_c}$ the $M_{\Phi_a}=130$\,GeV line in Fig.~\ref{figure:1} moves very
slowly to the left while, at the same time, $M_{\Sigma_3}=M_Z$ moves very
rapidly in the same direction until the triangular region shrinks to a point.
In other words, any scenario in which $M_{\Psi_T}$ or $M_{\Phi_c}$ or both are
below the GUT scale would be more significantly exposed to the tests through
the proton decay lifetime measurements and accelerator searches than the
scenario shown in Fig.~\ref{figure:1}.

If, on the other hand, one lowers the mass of $\Sigma_8$, the
$M_{\Phi_a}=130$~GeV line moves to the right more rapidly than the
$M_{\Sigma_3}=M_Z$ line until the triangular region becomes a point when
$M_{\Sigma_8}$ reaches $10^5$\,GeV. At that point $M_{GUT}$ reaches the upper
bound\footnote{We will confront this bound with the outcome of the two-loop
analysis and use it to evaluate the corresponding upper bound on the total
proton decay lifetime.} of $3.2 \times 10^{14}$~GeV for
$M_{\Sigma_8}=10^5$\,GeV, $M_{\Sigma_3}=M_Z$, $M_{\Phi_a}=130$\,GeV,
$M_{\Phi_b}=242$\,GeV and $\alpha_{GUT}^{-1}=37.3$. Again, the upper bound on
the masses of $\Sigma_3$, $\Phi_b$ and $\Phi_a$ would be significantly lower as
long as $M_{\Sigma_8} < M_{GUT}$.

\subsection{Two-Loop Analysis}

\begin{figure}[h]
\begin{center}
\includegraphics[width=4.5in]{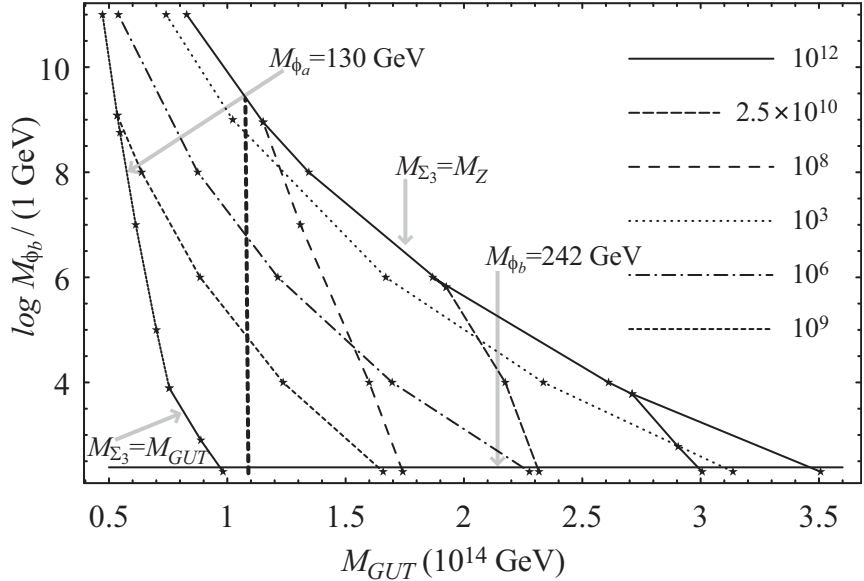}
\end{center}
\caption{\label{figure:2} Unification of the gauge couplings at the two-loop
level. Stars correspond to exact numerical two-loop unification solutions.
There are two sets of lines of constant value. The steeper set is associated
with $M_{\Phi_a}$ and the other one represents the lines of constant
$M_{\Sigma_3}$. All the masses are in GeV units. The region to the left of the
vertical dashed line is excluded by the proton decay experiments if
$\alpha=0.015$\,GeV$^3$.}
\end{figure}

The simplicity of the Higgs sector allows us to repeat the same analysis at the
two-loop level. We require exact unification and present the available
parameter space in Fig.~\ref{figure:2}. In the two-loop analysis we must also
take into account the one-loop running of the Yukawa couplings. The relevant
input parameters at the $M_Z$ scale, such as the fermion masses and CKM angles
that are used in the running are specified in Table I of
Ref.~\cite{Barr:2005ss}. The stars in Fig.~\ref{figure:2} represent points that
correspond to exact unification.

The upper bound on the GUT scale is shifted to a higher value (about a factor
of $\sqrt{2}$) with respect to the one-loop case. There also appears a line
along which $M_{\Sigma_3}$ is not required to be below $M_{GUT}$ to accomplish
exact unification. Moreover, the allowed values of $M_{\Phi_a}$ are
significantly larger than in the one-loop analysis. As before, by relaxing the
assumption $M_{\Psi_T},M_{\Phi_c}\geq M_{GUT}$ ($M_{\Sigma_3} \geq M_{GUT}$),
the allowed region moves to the left (right) and shrinks as we lower the
relevant masses.

At two-loop level, the corrected upper bound on $M_{GUT}$ is $4.6 \times
10^{14}$\,GeV for $M_{\Sigma_8}=M_Z$, $M_{\Sigma_3}=M_Z$, $M_{\Phi_a}=6.4
\times 10^3$\,GeV, $M_{\Phi_b}=242$\,GeV and $\alpha_{GUT}^{-1}=37.06$. We use
these values in the next section to derive an accurate upper bound on the
proton decay lifetime. If $\alpha=0.015$\,GeV$^3$ the proton decay lifetime
measurements already exclude the part to the left of the thick dashed line
($M_{GUT}\gtrsim 1.1 \times 10^{14}$\,GeV) and establish a triangular region
with the maximal value for $M_{\Phi_b}$ around $10^9$\,GeV.

With the two-loop analysis at hand, we can finally answer the following
question. How much improvement in the lifetime limits do proton decay
experiments need in order to completely exclude our scenario? In the ``worst''
case scenario the two-loop GUT scale is approximately by a factor of four
larger than the current proton decay bound presented in Fig.~\ref{figure:2}.
Hence, an improvement in the measurements of proton lifetime by a factor of
$4^4$ is called for to completely rule out this GUT scenario. The situation is
actually much better than that, since even the slightest improvement in the
proton lifetime bounds (by a factor of fifteen) will make our scenario
incompatible with exact unification unless either $\Phi_b$ or $\Sigma_3$
resides below $10^3$\,GeV, thus making them accessible in accelerator
experiments.

\subsection{$\Sigma \rightarrow
-\Sigma$ Invariance}

Fig.~\ref{figure:2} shows that $\Sigma_3$ and $\Sigma_8$ are highly
non-degenerate in some parts of the allowed parameter space. This, however,
seems in conflict with the tree-level relation $M_{\Sigma_3}= 4 M_{\Sigma_8}$
of the $\Sigma$ potential, which is invariant under the $\Sigma \rightarrow
-\Sigma$ transformation~\cite{Buras}. Since we commit to the scenario that
includes all possible terms allowed by the gauge symmetry, we are forced to
depart from this commonly used invariance. This has three important
consequences: (i) the cubic term ($\text{Tr}\, \Sigma^3$) in the potential
violates the validity of the $M_{\Sigma_3}= 4 M_{\Sigma_8}$
relation~\cite{Guth}; (ii) the higher dimensional terms linear in
$\Sigma/\Lambda$ in the Yukawa part of the Lagrangian allow for masses of
quarks and leptons that are in agreement with the experimentally observed
values~\footnote{Here $\Lambda$ is the scale where some new physics, relevant
for the ultraviolet (UV) completion of the theory, enters.}; (iii) a term
linear in both $\Sigma$ and $\bm{15}_H$ appears in the scalar part of the
potential that is relevant for both the proton decay and neutrino masses if
$\bm{15}_H$ couples to matter fields.

To demonstrate how constraining the demand for $\Sigma \rightarrow -\Sigma\,$
invariance is, we present one example in a somewhat simpler setting. Recall the
Georgi-Glashow $SU(5)$ model with two Higgs fields---one in the adjoint and the
other in the fundamental representation. After imposing the $\Sigma \rightarrow
-\Sigma$ invariance there are seven terms left in the scalar potential. It is
easy to show~\cite{Buras} that once $\Sigma_{24}$ gets a VEV $v_1$ of the order
of the GUT scale and $\Psi_D$ gets the electroweak scale VEV $v_2$, the
electrically neutral component of $\Sigma_{3}$ must get a VEV of the order of
$v_3 \sim v_2^2/v_1$ (For the relevant equations and further discussion
see~\cite{MaLi}. We note that there is a term $+\lambda_4 v_2 v_3^2$ missing on
the right hand side of Eq.~(11) of Ref.~\cite{MaLi}.) The above statement,
however, is not necessarily correct if we include two more terms in the
potential that are absent under the $\Sigma \rightarrow -\Sigma$ invariance. In
other words, a classical solution with appropriate VEVs for $\Sigma_{24}$ and
$\Psi_D$ but with no VEV for $\Sigma_{3}$ is allowed if the potential is not
invariant under $\Sigma \rightarrow -\Sigma\,$.

We conclude this section with the following observation. In our analysis we
include all the terms allowed by the underlying gauge symmetry. In this way we
insure that the predictions of our scenario are independent of any particular
set of assumptions. The benefit of such an approach is clear: if the
predictions of our scenario are experimentally refuted, then the scenario is
ruled out regardless of any particular assumptions.

\section{Upper bound on the total proton decay lifetime}
Proton decay is a generic prediction coming from matter unification. We thus
believe this to be the most promising way to test the beautiful idea of grand
unification. It is commonly thought that non-supersymmetric GUT scenarios are
ruled out by the limits on nucleon decay lifetimes, since the unification scale
is around $10^{14}$\,GeV. This, however, holds only in GUT scenarios where the
Yukawa sector is quite constrained. In general, this is no longer true because
proton decay predictions are different for each model of fermion
masses~\cite{Dorsner:2004xa}. To show that our minimal non-supersymmetric GUT
scenario based on $SU(5)$ gauge group is not ruled out by such limits, we look
for an upper bound on the total proton decay lifetime. For new experimental
lower bounds on the partial lifetime of the proton see Ref.~\cite{newbounds}.

It is well known that in any non-supersymmetric scenario the most important
contributions to the decay of the proton are the so-called gauge $d=6$
contributions. In the physical basis, these effective operators read
as~\cite{FileviezPerez:2004hn}:
\begin{eqnarray}
O( \nu_l, d_{\alpha}, d^C_{\beta}) &=& \frac{g_{GUT}^2}{2 M_V^2}
\ (V_1 \ V_{UD})^{1 \alpha} \ (V_3 \ V_{EN})^{\beta l} \ \overline{u^C}
\ \gamma^{\mu} \ L \ d_{\alpha}
\ \overline{d^C_{\beta}} \ \gamma_{\mu} \ L \ \nu_{l}\,, \\
O( e^C_{\alpha}, d_{\beta}) &=& \frac{g_{GUT}^2}{2 M_V^2} \
\left( V_1^{11} \ V_2^{\alpha \beta} \ + \ (V_1 \ V_{UD})^{1 \beta}
(V_2 \ V_{UD}^{\dagger})^{\alpha 1}\right)
\ \overline{u^C} \ \gamma^{\mu} \ L \ u \ \overline{e^C_{\alpha}}
\ \gamma_{\mu} \ L \ d_{\beta}\,, \\
O( e_{\alpha}, d^C_{\beta})&=& \frac{g_{GUT}^2}{2 M_V^2} \ V_1^{11}
 \ V_{3}^{\beta \alpha} \ \overline{u^C} \ \gamma^{\mu} \ L \ u \
\overline{d^C_{\beta}} \ \gamma_{\mu}\ L \ e_{\alpha}\,.
\end{eqnarray}
In the above equations $V_1= U_C^{\dagger}\, U$, $V_2=E_C^{\dagger}\,D$,
$V_3=D_C^{\dagger}\,E$, $V_4=D_C^{\dagger}\, D$, $V_{UD}=U^{\dagger}\,D$,
$V_{EN}=E^{\dagger}\,N$ are mixing matrices; $L=(1-\gamma_5)/2$ and
$\alpha,\beta=1,2$; $l=1,2,3$. Our convention for the diagonalization of the
up, down, charged lepton and neutral lepton Yukawa matrices is specified by
\begin{align}
U^T_C \ Y_U \ U = Y_U^{\textrm{diag}}\,,\quad D^T_C \ Y_D \ D =
Y_D^{\textrm{diag}}\,,\quad E^T_C \ Y_E \ E =Y_E^{\textrm{diag}}\,, \quad N^T \
Y_\nu \ N = Y_\nu^{\textrm{diag}}\,.
\end{align}
The quark and leptonic mixing are given in our notation by
$V_{UD}=U^{\dagger}D=K_1 V_{CKM} K_2$ and $V_{EN}=K_3 V_{PMNS}$, respectively,
where $K_1$, $K_3$ and $K_2$ are diagonal matrices containing three and two
phases, respectively.

The way to find an upper bound on the total proton decay lifetime by
investigating the possible freedom in the Yukawa sector of grand unified
theories has been pointed out in Ref.~\cite{Dorsner:2004xa}. For a given value
of $\alpha_{GUT}$ and the super-heavy gauge boson mass, it has been shown that
the upper bound in the case of Majorana neutrinos is given
by~\cite{Dorsner:2004xa}:
\begin{equation}
\tau_p \leq \frac{6}{\alpha_{GUT}^2} \times \left(\frac{M_V}{10^{16}
\textrm{GeV}}\right)^4 \times \left(\frac{0.003
\textrm{GeV}^3}{\alpha}\right)^2 \ \times 10^{39} \ \textrm{years}\,.
\end{equation}
Here, $\alpha$ is the value of the matrix element, usually taken
$\alpha=0.003\,\textrm{GeV}^3$ as a conservative value. However, in a recent
lattice calculation, $\alpha=0.015\,\textrm{GeV}^3$ has been
obtained~\cite{Aoki:2004xe}. (There is also a factor 4 difference between the
previous equation and Eq.~(9) of Ref.~\cite{Dorsner:2004xa}. This is due to an
erroneous normalization in~\cite{Dorsner:2004xa}.)

By inspecting the full parameter space where unification can be achieved in our
GUT scenario (cf. Figs.~\ref{figure:1} and \ref{figure:2}), it is then possible
to find the upper bound on the total proton decay lifetime. Using the maximal
$M_{GUT}$ value and associated values for $\alpha_{GUT}$, we can estimate the
upper bound on $\tau_p$, taking into account the one- and two-loop running of
the gauge couplings, respectively. Using
$\alpha=0.015\,\textrm{GeV}^3$~\cite{Aoki:2004xe}, these bounds read as
\begin{eqnarray} \label{taup_bounds}
\tau_p^{(\textrm{one-loop})}  \leq 3.5 \times 10^{35}\,\textrm{years},\\
\tau_p^{(\textrm{two-loop})}  \leq 1.4 \times
10^{36}\,\textrm{years}.
\end{eqnarray}
There is a difference of a factor 4 for the upper bounds on the total proton
lifetime between the two cases. As can be appreciated, our grand unified
scenario is not ruled out by the present experimental lower bound on the proton
decay lifetime (typically $\tau_{\rm exp} \gtrsim
10^{33}$~years~\cite{newbounds}). The regions that are ruled out are presented
by the vertical dashed thick lines in Figs.~\ref{figure:1} and~\ref{figure:2}.

In order to complete our study, let us also discuss the $d=6$ Higgs
contributions. In Ref.~\cite{Dorsner:2005fq}, the predictions coming from those
operators were studied in detail. It was shown that besides the usual $d=6$
Higgs terms there are also contributions due to the mixing between the colored
triplet $\Psi_T$ and the light leptoquark $\Phi_b$. This mixing comes from the
interaction term $c_3 \ 5_H^{\dagger} \ 15_H \ 5_H^*$. However, there is an
extra contribution to this mixing coming from the term $b_8 \ 5_H^{\dagger} \
24_H \ 15_H \ 5_H^*$ that we mentioned before. Therefore, when applying
Eq.~(10) of~\cite{Dorsner:2005fq} to the present case, one should replace $c_3$
by $c_3 - \lambda b_8/\sqrt{30} \equiv \tilde{c}_3/2$. Since the Higgs
contributions are very ambiguous, one can verify that the upper bound on the
proton decay lifetime in our GUT scenario is indeed given by
Eqs.~(\ref{taup_bounds}). In particular, we remark that in the present scenario
it is always possible to set to zero all Higgs contributions to the nucleon
decay by choosing the matrices $\underline{A}_{ij}=-\underline{A}_{ji}$ and
$D_{ij}=0$, except for $i=j=3$. (See Ref.~\cite{Dorsner:2005fq} for notation
and details.)

Certainly, if proton decay is not observed, the next generation of experiments
will improve the lower bounds on partial lifetimes by a few orders of
magnitude. For instance, the goal of Hyper-Kamiokande is to explore the proton
lifetime at least up to $\tau_p/B(p \to e^+ \pi^0) > 10^{35}$ years and
$\tau_p/B(p \to K^+ \bar{\nu})> 10^{34}$ years in about 10
years~\cite{Nakamura:2003hk}. Thus, our minimal GUT scenario will be tested or
ruled out at the next generation of proton decay experiments, since the
upper-bound on the total proton decay lifetime in our scenario is $\tau_p \leq
1.4 \times 10^{36}$ years.

\section{Constraints from triplet seesaw leptogenesis}

The origin of the baryon asymmetry observed in the universe is an outstanding
problem in particles physics and cosmology. The most recent Wilkinson Microwave
Anisotropy Probe (WMAP) results and big bang nucleosynthesis analysis of the
deuterium abundance imply~\cite{Spergel:2003cb}
\begin{equation}
\eta_{B}=\frac{n_B-n_{\bar{B}}}{n_\gamma}=(6.1\pm0.3)\times10^{-10}\,,\label{BAU}
\end{equation}
for the baryon-to-photon ratio of number densities. Among the viable mechanisms
to explain this primordial matter-antimatter asymmetry,
leptogenesis~\cite{Fukugita:1986hr} has undoubtedly become one of the most
compelling scenarios. Indeed, the evidence for non-vanishing neutrino masses
and the possibility that their origin is directly linked to lepton number
violation point towards leptogenesis as a natural mechanism for the generation
of the cosmological baryon asymmetry. Moreover, in GUT scenarios, where the
existence of heavy (boson or fermion) particles is predicted, leptogenesis can
be easily realized by means of the out-of-equilibrium decays of such particles
at temperatures below their mass scale. The lepton asymmetry generated in the
presence of $CP$-violating processes is then partially converted into a baryon
asymmetry by the sphalerons~\cite{Kuzmin:1985mm}.

In its simplest framework, consisting on the addition of hierarchical heavy
right-handed neutrinos to the standard model, successful leptogenesis implies a
lower bound on the mass of the lightest heavy Majorana neutrino, $M_{N_1}
\gtrsim 10^8\ (10^9)$~GeV, which holds assuming a thermal (zero) initial
abundance of the $N_1$ neutrinos before decaying. On the other hand, if the
same heavy neutrinos are responsible for the generation of the light neutrino
masses via the well-known seesaw mechanism~\cite{seesaw}, then their natural
mass scale is expected to be $M_N \sim v^2/m_\nu \sim 10^{14}$~GeV, for a light
neutrino mass scale around the atmospheric neutrino scale, i.e. $m_\nu \sim
m_{atm} \simeq 5\times 10^{-2}$~eV. However, in the presence of other
lepton-number violating interactions, such as the ones mediated by $SU(2)_L$
scalar triplets, the leptonic asymmetry produced by the out-of-equilibrium
decay of the heavy Majorana singlets can be totally washed out, if the triplet
mass scale is lower than $M_N$. In the latter case, leptogenesis could proceed
through the decay of the lightest triplet scalar. As we shall see below, the
$\Phi_a\, (\subset \bm{15}_H)$ $SU(2)_L$ scalar triplet with a mass $M_{\Phi_a}
\geq 10^{9-10}$~GeV constitutes a natural candidate for a successful triplet
seesaw leptogenesis in our minimal $SU(5)$ scenario.

To study the viability of thermal leptogenesis, we consider the simplest
extension of the minimal $SU(5)$ model proposed in~\cite{Dorsner:2005fq}, which
consists on the addition of right-handed neutrinos. We remark that, in the
present framework, the introduction of a single heavy Majorana neutrino is the
minimal extra particle content required to implement the leptogenesis mechanism
through the out-of-equilibrium decay of the triplet $\Phi_a$ into leptons
involving the virtual exchange of the right-handed neutrino. The relevant terms
of the right-handed Majorana neutrino and scalar triplet Lagrangian are
\begin{align}\label{Triplet_Lag}
    \mathcal{L} \owns -\frac{1}{2} M_{N} N^T C \, N - H^\dagger \bar{N}\, Y_N\, \ell
    -M_{\Phi_a}^2 \mathrm{Tr}\,\Phi_a^\dagger \Phi_a -\frac{1}{2} \ell^T C i
    \sigma_2 \Phi_a Y_\nu\, \ell + \frac{1}{2} \tilde{c}_3 H^T i \sigma_2 \Phi_a H +
    \mathrm{H. c.}\,,
\end{align}
with $\ell = (\nu, e)^T$, $H =(H^0, H^-)^T$,
\begin{align}\label{Triplet_def}
    \Phi_a = \begin{pmatrix}
\dfrac{1}{\sqrt{2}}\,\delta^+ & \delta^{++}  \\
\delta^0 & - \dfrac{1}{\sqrt{2}}\,\delta^+
\end{pmatrix} \,,
\end{align}
$M_N$ is the right-handed neutrino mass matrix, $Y_N$ and $Y_\nu$ are the
coupling matrices. For simplicity, flavour indices have been omitted. The
triplet (type-II seesaw) contribution to the effective neutrino mass matrix
$\mathcal{M}_\nu$ is given by
\begin{align}\label{Mnu_II}
    M_\nu^{II}=\tilde{c}_3 Y_\nu \frac{v^2}{M_{\Phi_a}^2}\,,
\end{align}
$v =\langle H_0 \rangle =174$~GeV, while the usual right-handed neutrino
(type-I seesaw) contribution is
\begin{align}\label{Mnu_I}
    M_\nu^I = -v^2 Y_N^T M_N^{-1} Y_N\,.
\end{align}

\begin{figure}[h]
  \includegraphics[width=12cm]{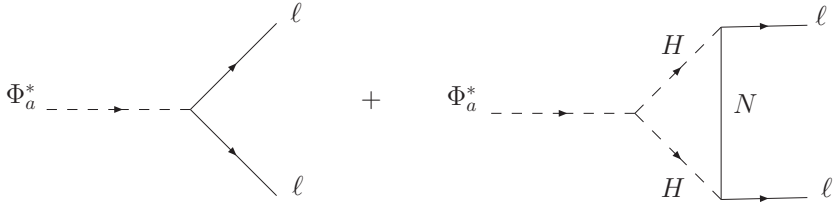}\\
  \caption{Tree-level and one-loop diagrams contributing to the $CP$ asymmetry in the
  scalar triplet decay.}\label{leptofig}
\end{figure}

In the presence of $CP$-violating interactions, the $\Phi_a$ decay into two
leptons generates a non-vanishing $CP$ asymmetry
\begin{align}
\varepsilon = 2\, {{\Gamma (\Phi_a^* \rightarrow \ell + \ell) - \Gamma (\Phi_a
\rightarrow \bar{\ell} + \bar{\ell })}\over{\Gamma_\Phi +\Gamma_{\Phi^*}}}\,,
\end{align}
where $\Gamma_\Phi$ denotes the total triplet decay width. Since in the present
minimal $SU(5)$ framework there are only two decay modes, $\Phi_a \rightarrow
\ell+\ell$ and $\Phi_a \rightarrow H+H$, one can write
\begin{align}\label{Decay_channels}
    \mathcal{B}_\ell\, \Gamma_\Phi  &\equiv  \Gamma(\Phi_a \rightarrow \ell+\ell)
    =\frac{M_{\Phi_a}}{16\pi} \text{Tr}\,
    Y_\nu^\dagger Y_\nu\,,\\
    \mathcal{B}_H \Gamma_\Phi &\equiv \Gamma(\Phi_a \rightarrow H+H)
    =\frac{1}{16\pi}\,\frac{|\tilde{c}_3|^2}{M_{\Phi_a}}\,,
\end{align}
where $\mathcal{B}_\ell$ and $\mathcal{B}_H$ are the corresponding tree-level
branching ratios ($\mathcal{B}_\ell+\mathcal{B}_H=1$). A nonzero $\varepsilon$
asymmetry is then generated by the interference of the tree-level decay process
with the one-loop vertex diagram, as shown in Fig.~\ref{leptofig}.

Assuming $M_{\Phi_a} \ll M_N$, the resulting asymmetry is approximately given
by~\cite{Hambye:2003ka}
\begin{align}\label{CPasym1}
   \varepsilon \simeq \frac{1}{8\pi^2}\,\frac{M_{\Phi_a}^3}{v^4}\,
   \frac{\mathrm{Im}\,[\mathrm{Tr}\, M_\nu^{II} M_\nu^{I^\dagger}
   ]}{\Gamma_\Phi}\,.
\end{align}
Using the relation
\begin{align} \label{GammaPhirel}
  16\pi v^2 \Gamma_\Phi \sqrt{\mathcal{B}_\ell\, \mathcal{B}_H} = M_{\Phi_a}^2
  \sqrt{ \mathrm{Tr}\,M_\nu^{II^\dagger} M_\nu^{II}}\,,
\end{align}
and recalling that the effective seesaw neutrino mass matrix is
$\mathcal{M}_\nu=M_\nu^{I}+M_\nu^{II}$, Eq.~(\ref{CPasym1}) can be recast in
the form~\cite{Hambye:2005tk}
\begin{align}\label{CPasym2}
   \varepsilon \simeq \frac{M_{\Phi_a} \sqrt{\mathcal{B}_\ell\, \mathcal{B}_H}}{4\pi v^2}\,
   \,
   \frac{\mathrm{Im}\,[\mathrm{Tr}\, M_\nu^{II} \mathcal{M}_\nu^{\dagger}
   ]}{\sqrt{\mathrm{Tr}\,M_\nu^{II^\dagger} M_\nu^{II}}}\,.
\end{align}

In order to discuss the bounds implied by leptogenesis, it is convenient to
write the $CP$ asymmetry $\varepsilon$ as the product
\begin{align}
\varepsilon =\varepsilon_{\rm max} \sin \delta_L\,,
\end{align}
where $\varepsilon_{\rm max}$ is the maximal $CP$ asymmetry and $\delta_L$ is
an effective leptogenesis phase. Using Eq.~(\ref{CPasym2}), it is then
straightforward to show that the following upper bound holds\footnote{This
value is typically much smaller than the maximal value allowed by unitarity, $
|\varepsilon| < 2 \min(\mathcal{B}_\ell, \mathcal{B}_H)$.}
\begin{align} \label{epsmax}
  \varepsilon_{\rm max} = \frac{M_{\Phi_a}}{4\pi v^2}\,
  \left(\mathcal{B}_\ell\, \mathcal{B}_H \sum m_{\nu_i}^2 \right)^{1/2}\,,
\end{align}
where $m_{\nu_i}$ are the light neutrino masses. Clearly, the absolute maximum
of the above expression is attained when $\mathcal{B}_\ell = \mathcal{B}_H
=1/2$. This situation, however, does not necessarily corresponds to a maximal
baryon asymmetry, since the efficiency of leptogenesis, dictated by the
solution of the relevant Boltzmann equations, is not necessarily maximal in
such a case. In fact, a numerical study of these equations shows that the
efficiency is minimal for $\mathcal{B}_\ell = \mathcal{B}_H =1/2$ and maximal
when either $\mathcal{B}_\ell \ll \mathcal{B}_H$ or $\mathcal{B}_\ell \gg
\mathcal{B}_H$~\cite{Hambye:2005tk}.

Assuming no pre-existing asymmetry, the total baryon asymmetry, obtained after
partial lepton-to-baryon conversion through the sphalerons, is given by
\begin{align} \label{etaB}
\eta_B \simeq - 3 \times 10^{-2}\,\varepsilon\,\kappa_f\,,
\end{align}
where $\kappa_f$ is the efficiency factor, normalized in such a way that
$\kappa_f$ approaches one in the limit of thermal initial $\Phi_a$ abundance
and no washout. The expression (\ref{etaB}), when combined with
Eq.~(\ref{epsmax}) leads to a lower bound on the mass $M_{\Phi_a}$. Indeed,
using the observed value $\eta_B^{\rm min} = 5.8 \times 10^{10}$ ( cf.
Eq.~(\ref{BAU})), one finds
\begin{align}\label{Mphilow}
M_{\Phi_a} \gtrsim \frac{1.5 \times 10^8~{\rm GeV}}{\kappa_f
\sqrt{\mathcal{B}_\ell\, \mathcal{B}_H}}\,\frac{0.05~{\rm eV}}{\sqrt{\sum
m_{\nu_i}^2}}\,.
\end{align}

It is possible to obtain a simple estimate of the efficiency of leptogenesis by
comparing the total triplet decay rate with the expansion rate of the universe.
We define, as usual, the decay parameter
\begin{align}\label{Kpar}
    K_\Phi \equiv \frac{\Gamma_\Phi}{H(T=M_{\Phi_a})}\,, \quad H(T) = 1.66\,
    g_\ast^{1/2} \frac{T^2}{M_P}\,,
\end{align}
where $H(T)$ is the Hubble rate, $M_P \simeq 1.2 \times 10^{19}$~GeV is the
Planck mass and $g_\ast$ is the effective number of relativistic degrees of
freedom ($g_\ast = 106.75$ in the SM). Using Eq.~(\ref{GammaPhirel}), we can
rewrite $K_\Phi$ in the form
\begin{align}
 K_\Phi =\frac{23}{\sqrt{\mathcal{B}_\ell\, \mathcal{B}_H}} \frac{\sqrt{\sum
m_{\nu_i}^2}}{0.05~{\rm eV}}\,.
\end{align}
For not very large values of $ K_\Phi \gtrsim 1$, an order-of-magnitude
estimate of the efficiency factor is $\kappa_f \sim 1/K_\Phi$ and
Eqs.~(\ref{Mphilow}) and (\ref{Kpar}) imply
\begin{align}
 M_{\Phi_a} \gtrsim \frac{3.4\times 10^9~{\rm GeV}}{\mathcal{B}_\ell\,
 \mathcal{B}_H}\,,
\end{align}
which for $\mathcal{B}_\ell = \mathcal{B}_H =1/2$ yields  $M_{\Phi_a} \gtrsim
1.4 \times 10^{10}$~GeV. Clearly, a more precise analysis requires the solution
of the full set of Boltzmann equations~\cite{Hambye:2005tk}. Nevertheless, for
$\kappa_f \simeq 10^{-2}-10^{-3}$ and $\mathcal{B}_\ell \sim \mathcal{B}_H$,
Eq.~(\ref{Mphilow}) implies
\begin{align}
 M_{\Phi_a} \gtrsim 3 \times 10^{10-11}~{\rm GeV} \left(\frac{0.05~{\rm eV}}{\sqrt{\sum
m_{\nu_i}^2}}\right)\,.
\end{align}
For hierarchical light neutrinos, one has $\sqrt{\sum m_{\nu_i}^2} \simeq
m_{atm}$ and the above bound leads to $M_{\Phi_a} \gtrsim 3 \times
10^{10-11}~{\rm GeV}$. On the other hand, if neutrinos are quasi-degenerate in
mass, then the WMAP constraint $\sum m_{\nu_i} < 0.69$~eV leads to the less
restrictive lower limit $M_{\Phi_a} \gtrsim 4 \times 10^{9-10}~{\rm GeV}$. We
also note that this bound could be further reduced by about a factor of two, if
the dark energy component of the universe is not in the form of a cosmological
constant. In the latter case, assuming a dark energy equation of state $p=-w
\rho$ with $w<-1$, the present cosmological bound on neutrino masses relaxes to
$\sum m_{\nu_i} < 1.48$~eV~\cite{Hannestad:2005gj}.

Combining the above leptogenesis bounds with the ones shown in
Fig.~\ref{figure:2}, we conclude that the natural implementation of the type-II
seesaw mechanism and successful leptogenesis exclude the region of the
parameter space where the mass of the triplet $M_{\Phi_a}$ is below $10^9 -
10^{10}$~GeV. This in turn implies that the leptoquark $\Phi_b$ must be light
enough ($M_{\Phi_b} \lesssim 10^{6-7}$~GeV) to satisfy the cosmological
constraints, thus opening the possibility to test our minimal
non-supersymmetric $SU(5)$ GUT scenario at the next generation of collider
experiments through the production of light leptoquarks, and particularly, at
LHC.

\section{Summary}
\label{conclusions}

We have investigated in detail the constraints coming from
unification of gauge interactions in the minimal extension of the
Georgi-Glashow model, where the Higgs sector is composed by
$\bm{5}_H$, $\bm{15}_H$ and $\bm{24}_H$. We have shown that the
scalar leptoquark $\Phi_b$ has to be light in order to achieve
unification in agreement with all experimental constraints. Using
the constraints coming from triplet seesaw leptogenesis, the upper
bound on the leptoquark mass is $M_{\Phi_b} \lesssim 10^{6-7}$
GeV. Therefore there is a hope that our scenario could be tested
at the next generation of collider experiments through the
production of these light leptoquarks.

We have also predicted an upper bound on the total proton decay lifetime which
is $\tau_p \leq \ 1.4 \times 10^{36}$ years. Since at the next generation of
proton decay experiments the bounds are expected to be improved by a few orders
of magnitude, this minimal non-supersymmetric $SU(5)$ model will be certainly
tested or ruled out.

The upper bound on the proton decay lifetime and the exciting possibility to
verify the model at future collider experiments make our GUT scenario an
appealing candidate for the testability of the idea of grand unification.

\begin{acknowledgments}
The work of P.F.P. has been supported by {\em Funda\c{c}\~{a}o para a Ci\^{e}ncia e a
Tecnologia} (FCT, Portugal) through the project CFTP, POCTI-SFA-2-777 and a
fellowship under project POCTI/FNU/44409/2002. The work of R.G.F. was supported
by FCT under the grant SFRH/BPD/1549/2000.
\end{acknowledgments}

\appendix
\section{Two-loop gauge coupling running}
\label{first} The relevant two-loop equations for the running of the gauge
couplings take the form
\begin{equation}
\mu \frac{\textrm{d}\,\alpha_i(\mu)}{\textrm{d}\,\mu} =\frac{b_i}{2 \pi}\,
\alpha^2_i(\mu)+\frac{1}{8 \pi^2} \sum_{j=1}^{3} b_{ij}\,
\alpha^2_i(\mu)\,\alpha_j(\mu)+\frac{1}{32 \pi^3}\, \alpha^2_i(\mu)
\sum_{l=U,D,E} \text{Tr}\,[C_{il} Y^\dagger_l Y_l]\,.
\end{equation}
The general formula for $b_i$ and $b_{ij}$ coefficients is given
in~\cite{Jones:1981we}. Besides the well-known SM coefficients we have:
\begin{equation*}
b_{i}^{\Sigma_8(\Sigma_3)}= \left(\begin{array}{c}
   0 \\
   0\,(\frac{1}{3}) \\
   \frac{1}{2}\, (0)\\
\end{array}\right),
\qquad b_{i}^{\Phi_b}= \left(\begin{array}{c}
   \frac{1}{30} \\
   \frac{1}{2} \\
   \frac{1}{3} \\
\end{array}\right),
\qquad b_{i}^{\Phi_a}= \left(\begin{array}{c}
   \frac{3}{5} \\
   \frac{2}{3} \\
   0 \\
\end{array}\right),
\end{equation*}
\begin{equation*}
b_{ij}^{\Sigma_8(\Sigma_3)}=\left(
\begin{array}{ccc}
  0 & 0 & 0 \\
  0 & \quad 0\,(\frac{28}{3}) & 0 \\
  0 & 0 & 21\,(0) \\
\end{array}\right),
\quad b_{ij}^{\Phi_b}=\left(
\begin{array}{ccc}
  \frac{1}{150} & \frac{3}{10} & \frac{8}{15} \\
  \frac{1}{10} & \frac{13}{2} & 8 \\
  \frac{1}{15} & 3 & \frac{22}{3} \\
\end{array}\right),
\quad b_{ij}^{\Phi_a}=\left(
\begin{array}{ccc}
  \frac{108}{25} & \frac{72}{5} & 0 \\
  \frac{24}{5} & \frac{56}{3} & 0 \\
  0 & 0 & 0 \\
\end{array}\right),
\end{equation*}
which we incorporate at the appropriate scales. The $C_{il}$ coefficients
are~\cite{Arason:1991ic}:
\begin{equation*}
C_{il}=\left(
\begin{array}{ccc}
  \frac{17}{10} & \frac{1}{2} & \frac{3}{2} \\
  \frac{3}{2} & \frac{3}{2} & \frac{1}{2} \\
  2 & 2 & 0 \\
\end{array}\right).
\end{equation*}

To insure the proper inclusion of boundary conditions~\cite{Hall:1980kf} at
$M_{GUT}$ we set
$\left.\alpha^{-1}_i\right|_{GUT}=\alpha^{-1}_{GUT}-\lambda_i/(12 \pi)$, where
$\{\lambda_1,\lambda_2,\lambda_3\}=\{5,3,2\}$. The one-loop equations for the
Yukawa couplings can be found, for example, in Ref.~\cite{Arason:1991ic}.

\newpage


\end{document}